\documentclass[aps,prl,twocolumn,showpacs,superscriptaddress]{revtex4-2}
\usepackage{graphicx,amsmath,amssymb,amsfonts}
\usepackage{array}
\usepackage{multirow}
\usepackage{float}
\usepackage{color}
\usepackage[normalem]{ulem}
\usepackage{booktabs}
\usepackage{siunitx}
\usepackage{mhchem}
\usepackage{hyperref}
\hypersetup{
	colorlinks = true,
	linkcolor = [rgb]{0.70, 0.13, 0.13},
	citecolor = [rgb]{0.25, 0.41, 0.88},
	urlcolor = [rgb]{0.25, 0.41, 0.88}
}
\usepackage{pifont}

\begin{document}

\title{Noncollinear antiferromagnetic structure and physical properties of CrRhAs with distorted kagome lattice}

\author{Chenglin Shang}
\author{Daye Xu}
\author{Bingxian Shi}
\author{Xuejuan Gui}
\author{Zhongcen Sun}
\author{Juanjuan Liu}
\author{Jinchen Wang}
\author{Hongxia Zhang}
\affiliation{Laboratory for Neutron Scattering and Key Laboratory of Quantum State Construction and Manipulation (Ministry of Education), School of Physics, Renmin University of China, Beijing 100872, China}

\author{Hongliang Wang}
\author{Lijie Hao}
\affiliation{China Institute of Atomic Energy, PO Box-275-30, Beijing 102413, China}
\author{Peng Cheng}
\email[Corresponding author: ]{pcheng@ruc.edu.cn}
\affiliation{Laboratory for Neutron Scattering and Key Laboratory of Quantum State Construction and Manipulation (Ministry of Education), School of Physics, Renmin University of China, Beijing 100872, China}

\begin{abstract}
	
CrRhAs was theoretically proposed to be a kagome metal with unusual magnetic ground states, however little is known about its magnetic structure and physical properties experimentally. Here, we present an experimental investigation on CrRhAs with ZrNiAl-type structure and distorted Cr-kagome lattice. CrRhAs is an antiferromagnet with $T_N$ = 149~K. Powder neutron diffraction analysis reveals a
noncollinear antiferromagnetic structure with propagation vector $k=(1/3,1/3,1/2)$, which features a ferromagnetic second nearest neighbor coupling in the kagome plane that is different from the prediction in previous density functional theory calculations. Furthermore, CrRhAs exhibits anomalous electrical transport properties which are possibly related to multi-band effects and strong spin fluctuations. For the temperature-dependent longitudinal resistivity $\rho_{xx}$, it is semiconducting-like above $T_N$ and becomes metallic below $T_N$. The Hall coefficients exhibit two sign changes near 70~K and 300~K. Combined with the results of heat capacity measurements, a large Kadowaki-Woods ratio $\alpha=33.9$ $\mu$$\Omega$~cm~mol$^2$~K$^2$/J$^{2}$ is obtained. The above results suggest CrRhAs is a strongly correlated kagome metal with multi-band and noncollinear magnetic structure features. 

\end{abstract}

\maketitle

\section{Introduction}
Kagome lattice may naturally host both geometric spin frustration and topological band structures with the coexistence of flat-bands, Dirac crossings and Van Hove singularities\cite{Yin2022}. Therefore it has become one of the most fascinating crystal model in condensed matter physics for decades. Recently, various emergent quantum phenomena such as superconductivity\cite{AV3Sb5_Superconductivity_NRM}, charge-density-waves\cite{CDW1,CDW2,ScV6Sn6,FeGe}, unconventional magnetism\cite{TbMn6Sn6,FBFM,FBFM2018} and topological Hall effects\cite{Mn3Sn_AnomalousHall_N,YMn6Sn6_TopologicalHall_PRB,Fe3Sn2_TopologicalHall_PRL} has been discovered in metallic kagome material which makes it a fertile ground to study the interplay between geometry, topology, and electronic correlations.  

Perfect kagome lattice consists of coplanar equilateral triangles and regular hexagons with the same side length. It is known that many intriguing phenomena related to kagome physics do not necessarily require a perfect kagome lattice. For examples, Fe$_3$Sn$_2$ has flat-bands near Fermi level which possibly results in its high-temperature ferromagnetic ordering\cite{FBFM2018,Fe3Sn2_FlatBand_PRB}. Mn$_3$Sn is a
magnetic Weyl semimetal with topological Hall effect due to the chiral spin texture\cite{Mn3Sn_WeylFermions_NM,Mn3Sn_AnomalousHall_N}. Both Fe$_3$Sn$_2$ and Mn$_3$Sn have a slightly in-plane distorted kagome lattice. Furthermore, recent theoretical calculations have shown that even for Fe$_2$P-type materials with strongly distorted kagome lattice, topological band features with nodal-ring band crossings and Weyl points may still exist\cite{Kagome_Physics_AM}.

Fe$_2$P- or ZrNiAl-type intermetallic compounds belong to a large family of kagome metals\cite{ZrNiAl}. Although their kagome lattice is notably twisted by triangle rotations, the basic kagome geometry is preserved in them. HoAgGe is one of the ZrNiAl-type intermetallics with Ho-sites in the $ab$-plane forming the twisted kagome lattice. A recent neutron scattering investigation on HoAgGe has discovered that its local easy-axis anisotropy together with ferromagnetic nearest-neighbour coupling of the Ho$^{3+}$ moments lead to an exotic kagome spin ice state, which obeys two-in-one-out or one-in-two-out local arrangements of the spins on the lattice-triangles ("ice rule")\cite{HoAgGe_SpinIce_S}. Further study reveals that partially due to the non-trivial distortion of the kagome lattice, an emergent time-reversal-like degeneracy in HoAgGe results in contrasting hysteretic behavior of magnetization and magnetotransport data\cite{HoAgGe_DiscreteDegeneracies_NP}. These discoveries have drawn considerable research interests in this family of kagome metals.

Although there have been many studies on the ZrNiAl-type kagome metals with rare-earth ions constituting the kagome lattice\cite{HoAgGe_SpinIce_S,HoAgGe_DiscreteDegeneracies_NP,GdPtPb_PRB,HoPtSn_FirstOrderMagneticTransition_PRM,LuRuGe_HallCoefficientChange_IC,TbPtIn_AngularDependentMetamagnetism_PRB,YbAgGe_QuantumBicriticality_PRL,TmPtIn_BGM_JMM}, the experimental researches on this type of material with kagome lattice consisting of $3d$ transitional-metal ions are very rare\cite{MnRuP_HelicalMagneticOrder_PRR,MnRhAs_MagnetocaloricEffects_JAP}. CrRhAs is one member with Cr-based twisted kagome lattice similar as that of HoAgGe. Recently, density functional theory (DFT) calculations show that CrRhAs has a noncollinear ground state magnetic structure results from an unusual magnetic Hamiltonian and strong antiferromagnetic second nearest-neighbor coupling\cite{CrRhAs_Theory_NPJ}. However, little is known about the physical properties of CrRhAs experimentally besides an antiferromagnetic transition near 165~K identified several decades ago\cite{CrRhAs_Magnetic_JMM}. Therefore it is important to check the theoretical predicted magnetic structure of CrRhAs experimentally. In addition, it is also quite interesting for a deep investigation on its magnetic and transport properties.

In this study, the fabrications of both polycrystalline samples and micro-single crystals of CrRhAs are reported. X-ray photoelectron spectroscopy, neutron diffraction, magnetic susceptibility, electrical transport and specific heat measurements have been performed on CrRhAs with several intriguing findings. Noncollinear antiferromagnetic structure with in-plane magnetic anisotropy is identified below T$_N$ $\sim$ 149~K, which has some major differences with previous DFT calculated one. Sign reversals of Hall coefficients and other transport behavior suggest CrRhAs is a multi-band metal with strong spin fluctuations above T$_N$. A large Kadowaki-Woods ratio is derived implying the existence of strong electron correlations in this compound. These features make CrRhAs a special member in the ZrNiAl-type material family.

\section{methods}
Polycrystalline samples of CrRhAs were synthesized by solid-state reaction method. Firstly, stoichiometric Cr, Rh and As powders were ground and pressed into a pellet. Then it was placed into an alumina crucible and sealed in an evacuated quartz tube. The tube was firstly heated to \SI{500}{\celsius} and maintained for two days then to \SI{800}{\celsius} and kept for two days. After furnace cooled to room temperature, the product was reground, pressed and sealed in an evacuated quartz tube again. A final sintering was made at \SI{900}{\celsius} for three days and phase-pure polycrystalline CrRhAs could be obtained.

Several kinds of metal flux have been tried to grow single crystals of CrRhAs and only the Pb-flux method is successful. Firstly, Cr, Rh, As and Pb powders were mixed with a molar ratio of 2.5:1:1:18 and placed in an alumina crucible, then sealed in an evacuated quartz tube. The tube was heated up to \SI{1100}{\celsius} then maintained at this temperature for one day. Then it was slowly cooled to \SI{700}{\celsius} at a rate of \SI{1}{\celsius}/h. The CrRhAs single crystals could be found at the bottom of the alumina crucibles after centrifuging the Pb flux at this temperature. These rod-like crystals have a typical size of 0.5 $\times$ 0.05 $\times$ 0.05 mm$^{3}$, which are not large enough for most physical property measurements. Therefore most of the experimental measurements in this work were done on polycrystalline samples.

\begin{figure*}[htbp]
	\centering
	\includegraphics[width=\textwidth]{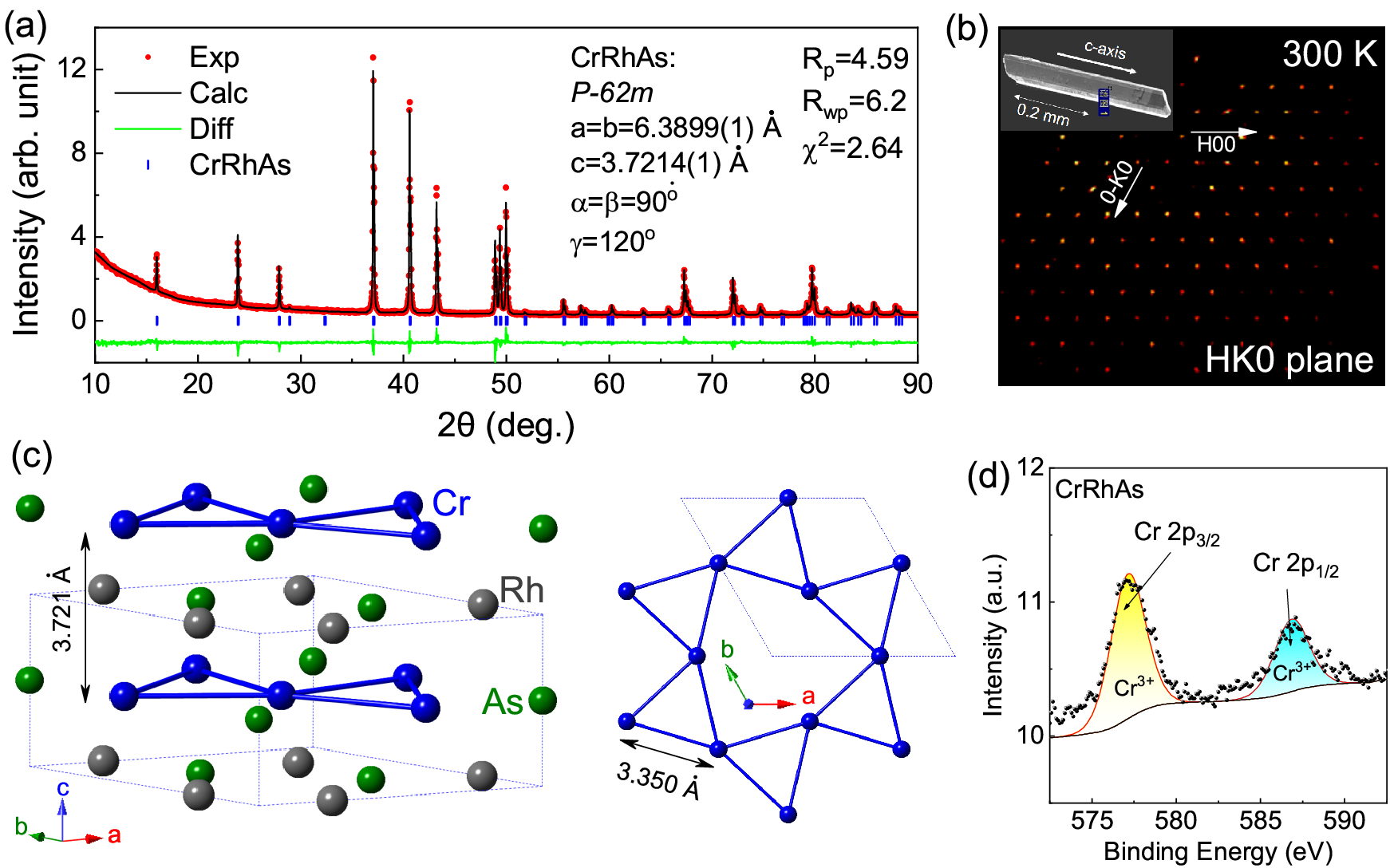}
	\caption{Structural characterization of CrRhAs. (a) XRD patterns of CrRhAs powder samples and the Rietveld refinement result. (b) Precession photo of (HK0) plane obtained from single crystal XRD for a CrRhAs single crystal. The inset shows a SEM photo for a typical CrRhAs single crystal. (c) A sketch of the crystal structure of CrRhAs. The Cr-based distorted kagome lattice viewed along $c$-axis is shown in the right side. The crystal unit cell is illustrated using blue dotted line. (d) The XPS spectra of Cr $2p$ core levels for CrRhAs.} \label{Fig1}
\end{figure*}

\begin{table*}
	\centering
	\caption{Atomic coordinates and occupancy obtained from
		the Rietveld refinement of powder XRD data.}
	\begin{tabular}{ccccccc}
		\toprule
		atom & Wyckoff & occupancy & x & y & z  \\
		\midrule
		Rh & 3f	 & 1 & 0.2568(3) & 0 & 0 \\
		Cr & 3g & 1 & 0.5911(5) & 0 & 0.5 \\
		As1 & 1b & 1 & 0 & 0 & 0.5 \\
		As2 & 2c & 1 & $1/3$ & $2/3$ & 0 \\
		\bottomrule
	\end{tabular}
	\label{1}
\end{table*}

X-ray diffraction (XRD) data of the CrRhAs single crystals and powders were collected at room temperature from a Bruker D8 VENTURE PHOTO II diffractometer equipped with multilayer mirror monochromatized Mo K$_{\alpha}$ ( ${\lambda}$ = 0.71073 {\AA} ) radiation and Bruker D8 Advance X-ray diffractometer using Cu K$_{\alpha}$ radiation, respectively. The single crystal Precession photos were obtained using the APEX3 program. The elemental composition of single crystals were examined with energy dispersive x-ray spectroscopy (EDS, Oxford X-Max 50). The valence states of the sample were examined using X-ray photoelectron spectroscopy (XPS, Thermo Fisher K-Alpha). Magnetization data were collected on a Quantum Design Magnetic Property Measurement System (MPMS). The electrical transport and specific heat measurements were carried out on a Quantum Design Physical Property Measurements System (PPMS-14T). The powder neutron diffraction experiments were carried out on Xingzhi cold neutron triple-axis spectrometer at the China Advanced Research Reactor (CARR)\cite{XingZhi}. About 8 grams of CrRhAs powder were used in the neutron experiments and the incident neutron energy was fixed at 16~meV.

\section{Results and discussions}

\subsection{Crystal structure and magnetization}

The ZrNiAl-type crystal structure of CrRhAs was examined by powder XRD. It crystallizes in the noncentrosymmetric hexagonal space group $P\overline 6 2m$ (No.189) with the refined crystallographic data shown in the inset of Fig.\ref{Fig1}(a) and Table.1, similar to that in previous reports\cite{CrRhAs_Magnetic_JMM,CrRhAs_Magnetic_JAP}. For the small single crystals grown from Pb-flux, both EDS and single crystal XRD in Fig.\ref{Fig1}(b) confirm they belong to the same CrRhAs phase. These crystals are rod-like that the long axis is the crystallographic $c$-axis as shown in the inset of Fig.\ref{Fig1}(b). Fig.\ref{Fig1}(c) illustrates the crystal structure of CrRhAs. The magnetic ions Cr form a distorted kagome lattice same as that in kagome spin ice compound HoAgGe\cite{HoAgGe_SpinIce_S}. Although it has a noticeable twist by triangle rotations comparing with a perfect kagome lattice, the kagome connectivity remains in this type of structure that may realize most of the kagome physics\cite{CrRhAs_Theory_NPJ}. This distorted Cr kagome lattice is also quasi-two-dimensional as the interlayer Cr-Cr distance (3.721 \AA) is larger than the intralayer Cr-Cr distance (3.350 \AA). Fig.\ref{Fig1}(d) displays the XPS spectra of Cr $2p$ core levels in CrRhAs. The peaks related to Cr $2p_{3/2}$ and $2p_{1/2}$ core levels are observed. Their binding energies reveal a Cr$^{3+}$ valence state in this compound, which is consistent with the following analysis on the magnetization data.

\begin{figure*}[htbp]
	\centering
	\includegraphics[width=\textwidth]{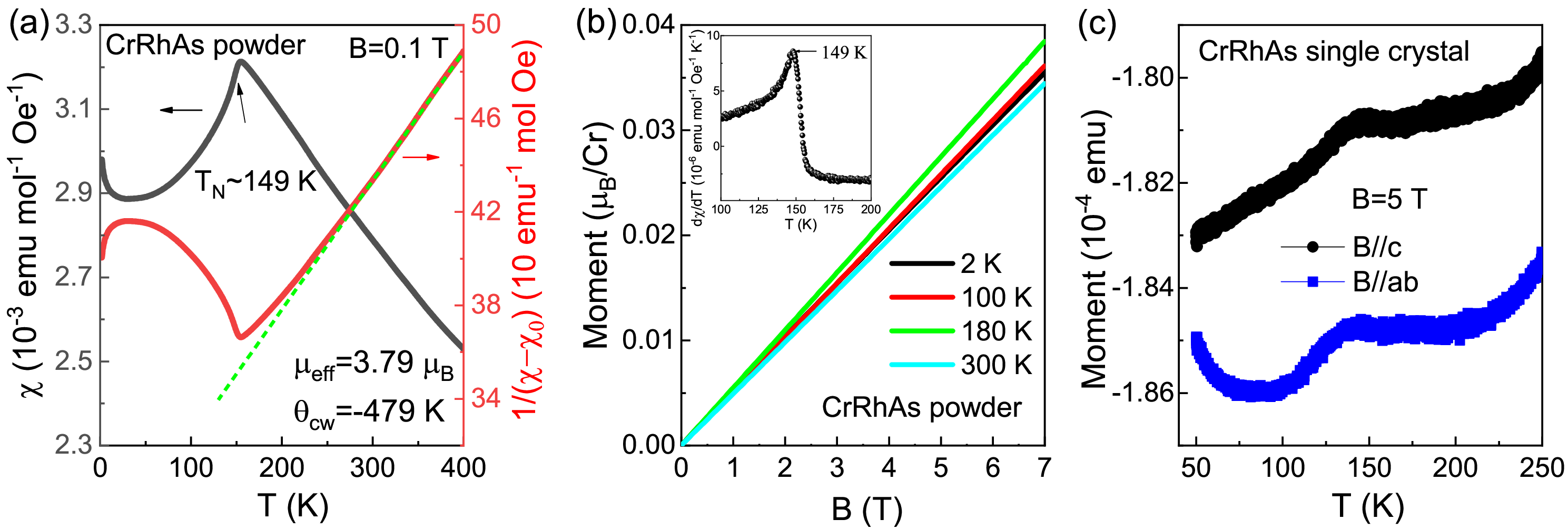}
	\caption{Magnetization measurements on CrRhAs. (a) Temperature dependence of magnetic susceptibility under magnetic field of 0.1~T. The green dashed line is a Curie-Weiss fit on the inverse susceptibility above 300~K. (b) Isothermal magnetization measured at different temperatures. The inset shows the d$\chi$(T)/dT curve. (c) Raw data of magnetization measured on a CrRhAs single crystal with magnetic field of 5~T applied parallel and perpendicular to the $c$-axis.} \label{Fig2}
\end{figure*}

The temperature dependent magnetic susceptibility $\chi$(T) of CrRhAs powder samples is plotted Fig.\ref{Fig2}(a). An antiferromagnetic transition is identified at $T_N$$\sim$149~K. This value is determined by the peak temperature in the d$\chi$/dT curve shown in the inset of Fig.\ref{Fig2}(b) and slightly lower than that in early reports\cite{CrRhAs_Magnetic_JMM,CrRhAs_Magnetic_JAP}. From above $T_N$ to room temperature, $\chi$(T) does not follow a well defined Curie-Weiss (CW) paramagnetic behavior. Previous works performed CW fit on this temperature range and gave effective moment and Weiss temperature significantly lower than that revealed from experimental fact and theoretical calculations\cite{CrRhAs_Magnetic_JMM,CrRhAs_Theory_NPJ}. In this work, $\chi$(T) is measured up to 400~K as shown in Fig.\ref{Fig2}(a). CW fit using equation $\chi=\chi_0+1/(T-\theta_{CW})$ on $\chi$(T) data from 300~K to 400~K yields effective moment $\mu_{eff}=3.79~\mu_{B}/Cr$ and CW temperature $\theta_{CW}=-479$~K. The obtained effective moment agrees well with the $S=3/2$ spin state for Cr$^{3+}$ that determined from XPS spectra. The CW temperature reveals strong antiferromagnetic interactions and the value is quite close to $\theta_{CW}=-578$~K given in previous theoretical calculations\cite{CrRhAs_Theory_NPJ}. The above result also reveals certain spin frustration effect ($\theta_{CW}$ : $T_N$$\sim$ 3.2) in this compound. In addition, the deviation of CW behavior in $\chi$(T) curve from above $T_N$ to 300~K suggests short-range magnetic order or spin fluctuations could exist, which may possibly be responsible for the anomalous transport behavior shown later.     

Fig.\ref{Fig2}(b) shows the isothermal magnetization data M(B) on powder samples. For different temperatures either below or above $T_N$, M(B) curves are linear and the difference of their slopes is very small, consistent with corresponding antiferromagnetic or paramagnetic states. Temperature dependent magnetization measurements were also carried out on a single crystal of CrRhAs and the raw data is shown in Fig.\ref{Fig2}(c). Because the mass of these micro-single crystals are all very small, the magnetization signal from CrRhAs crystal is accompanied by the diamagnetic background from the sample holder. However the antiferromagnetic transition is still visible near $T_N$ when applying a field of 5~T. Comparing the data under different field directions, one can see the susceptibility drop below $T_N$ is a bit sharper for B$\parallel$ab which implies an easy-plane magnetic anisotropy that is consistent with the result given by the neutron diffraction experiment.

\subsection{Magnetic structure}

\begin{figure*}[htbp]
	\centering
	\includegraphics[width=\textwidth]{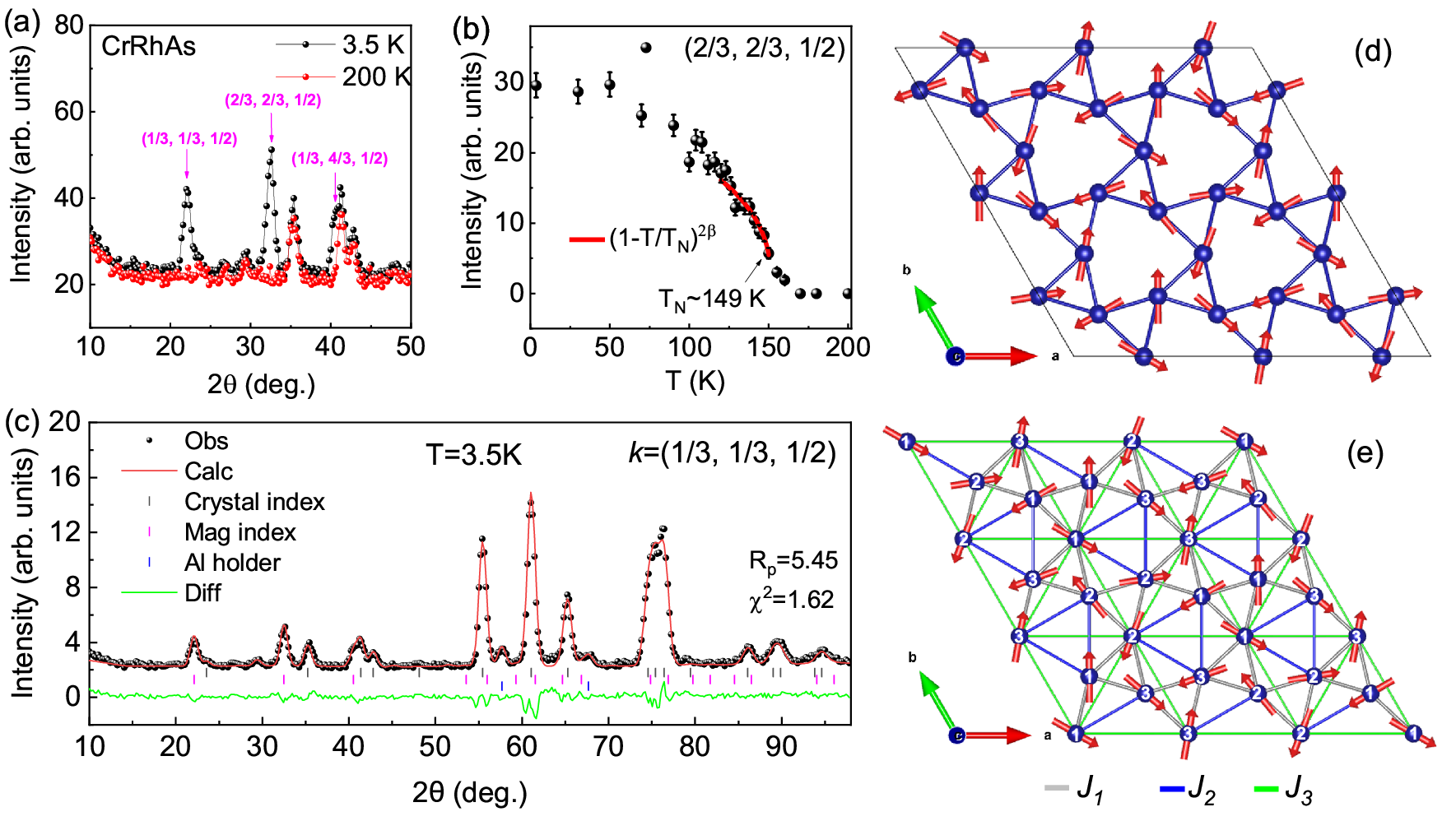}
	\caption {Magnetic structure investigated by neutron diffraction. (a) Powder neutron diffraction patterns of CrRhAs at 3.5~K and 200~K in the low-angle range. The observed magnetic Bragg peaks are indexed. (b) Measurements of the magnetic order parameter upon warming at the (2/3, 2/3, 1/2) reflection. (c) Rietveld refinement result on the neutron diffraction data at 3.5~K based on the magnetic structure plotted in (d) and (e). (d,e) Illustration of the low-temperature magnetic structure of CrRhAs in one kagome layer. Only the nearest Cr-Cr bondings are connected for (d), while for (e) the intraplane nearest ($J_1$), second nearest ($J_2$) and third nearest ($J_3$) bondings are connected using different colored solid lines. The number labels in (e) indicate the three inequivalent positions of Cr according to the magnetic space group.} \label{Fig3}
\end{figure*}

Neutron diffraction experiments were performed on CrRhAs powder to explore its antiferromagnetic structure. Fig.\ref{Fig3}(a) shows the data at 3.5~K and 200~K in the low-$Q$ region. The diffraction patterns at 50~K were also collected but almost overlap with that at 3.5~K within the instrumental resolution. Three magnetic Bragg peaks could be clearly observed and indexed as (1/3, 1/3, 1/2), (2/3, 2/3, 1/2) and (1/3, 4/3, 1/2) respectively. The temperature dependent integrated intensities of the (2/3, 2/3, 1/2) magnetic peak are shown in Fig.\ref{Fig3}(b). Fitting the temperature dependence of these integrated peak intensities near T$_N$ with a power law term $(1-T/T_N)^{2\beta}$ gives T$_N$ = 151(1)~K and $\beta$ = 0.185(1). Comparing with that in Ising, XY, Heisenberg or other magnetic models, the value of critical component $\beta$ is more closer to 0.125, which is expected for a two-dimensional Ising model. This possibly suggests that CrRhAs might belong to a quasi-two-dimensional system, which is consistent with the theoretically proposed weak interlayer magnetic exchange interaction\cite{CrRhAs_Theory_NPJ}. Furthermore, there is a magnetic intensity tail	extending to $\sim$160~K for (2/3, 2/3, 1/2). For one thing, it should be the magnetic contribution at the onset Neel temperature $\sim$155~K determined by the peak of $\chi(T)$ curve. For another, it is also consistent with the short-range magnetic order or strong spin fluctuations above $T_N$ that inferred from magnetization data\cite{Ba122}. 

All the magnetic Bragg peaks could be well indexed by a propagation vector $k$ = (1/3, 1/3, 1/2). According to the symmetry analysis done by the Bilbao crystallographic server\cite{Bilbao}, for the parent space group $P\overline 6 2m$ and this propagation vector, there are three possible magnetic subgroups that are: $P_c \overline 6, P_c \overline 6 c2, P_c \overline 6 m2$. The first two have either in-plane or out-of-plane basic vectors, the last one has only in-plane basic vectors. Among all these five possible magnetic models, the neutron diffraction data at 3.5~K can only be well fitted by the magnetic structure with magnetic space group $P_c \overline 6$ which has only in-plane basic vectors. Its refinement result and goodness-of-fit parameters are shown in Fig.\ref{Fig3}(c). The other four possible magnetic structures cannot give a fit to the observed magnetic Bragg peaks with acceptable evaluation factors. Especially for the models with apparent in-plane antiferromagnetic next-nearest-neighbor coupling, their calculated intensities of the major magnetic Bragg peaks are far below that observed experimentally. It should be mentioned that the magnetic unit cell is eighteen times larger than the nuclear unit cell. In the magnetic unit cell with space group $P_c \overline 6$, the magnetic ions Cr has three inequivalent positions indicated by the number labels in Fig. 3(e). Initial refinement without any constraint would result in different effective total moments for different Cr ions. For Cr ions in CrRhAs with only one Wyckhoff position in the paramagnetic nuclear cell, it is more physically reasonable that all Cr ions have the same magnitude of magnetic moment, as in previous theoretical calculations\cite{CrRhAs_Theory_NPJ}. Therefore this constraint is added for the Rietveld refinement shown in Fig.\ref{Fig3}(c) to obtain the final magnetic structure. It also should be noted that although the magnitude of magnetic moments and the in-plane tilting angle of the spins would both slightly vary without this constraint, the main features of the magnetic structure described below would remain the same.

Fig. \ref{Fig3}(d) and (e) show the refined magnetic structure of CrRhAs. The magnetic unit cell and plot style chosen for (d) is the same as that in previous report of HoAgGe for a clear comparison with that of a kagome spin ice state\cite{HoAgGe_SpinIce_S}. While the magnetic unit cell shown in (e) and the connections of different Cr-Cr bondings ($J_1$, $J_2$ and $J_3$) are similar as that in previous DFT calculations of CrRhAs\cite{CrRhAs_Theory_NPJ}, which makes the comparison with theoretical proposed magnetic structure easier. First of all, the ordered moments of Cr lie in the $ab$-plane without any component along the $c$-axis. This easy plane magnetic anisotropy is consistent with the magnetization data and previous theoretical calculations\cite{CrRhAs_Theory_NPJ}. The size of effective moment per Cr is $2.21\pm0.04~\mu_{B}$, which is lower than the theoretical value $3~\mu_{B}$ by assuming $g$=2 in a localized model. The reduced moment value is possibly due to itinerant magnetism may exist in CrRhAs since it is a metallic system. Secondly, the magnetic structure shows a  noncollinear in-plane configuration and the interlayer coupling is strictly antiferromagnetic. Only the spin configuration of one kagome layer is shown in Fig.\ref{Fig3} and one can obtain the magnetic structure of the adjacent kagome layer by reversing the directions of all spins. The features described above agree well with that in previous DFT calculations\cite{CrRhAs_Theory_NPJ}. Thirdly, in contrast to the prediction of a strong antiferromagnetic second nearest neighbor coupling ($J_2$) in the kagome plane, our result actually gives a ferromagnetic spin configuration for Cr ions connected by $J_2$ as shown by the blue triangles in Fig.\ref{Fig3}(e). Further discussions will be given in the DISCUSSION subsection.

\begin{figure*}[htbp]
	\centering
	\includegraphics[width=\textwidth]{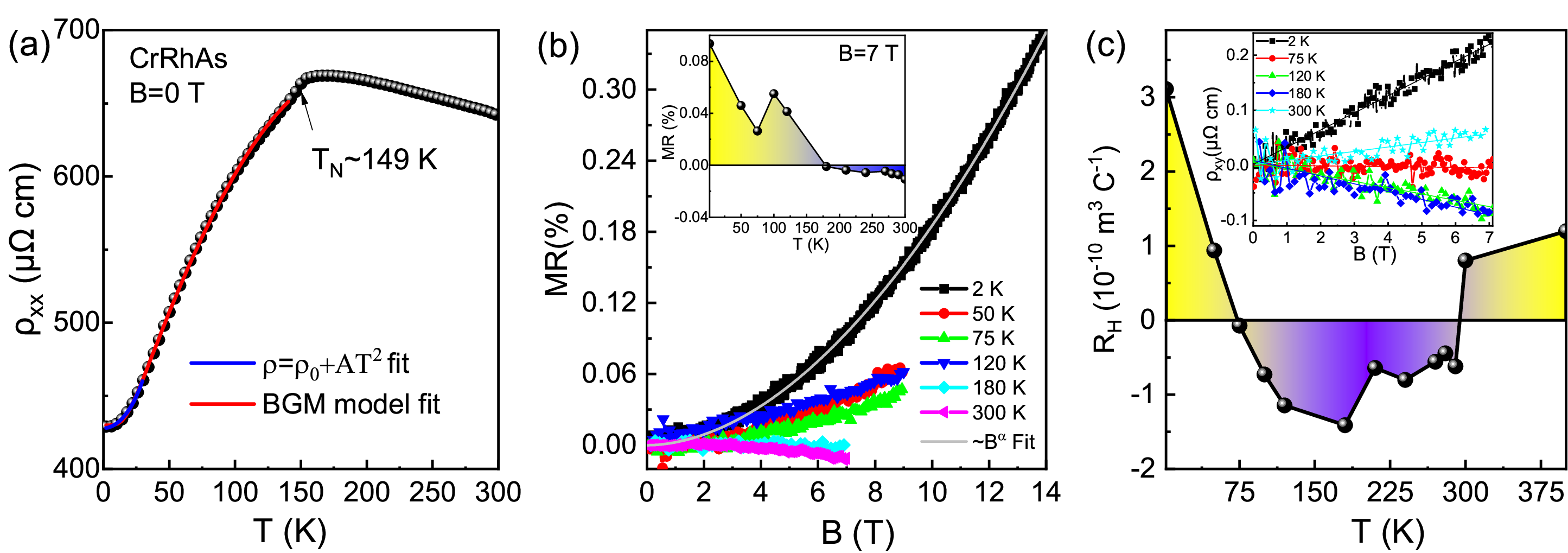}
	\caption {Electrical transport properties. (a) Resistivity of CrRhAs as a function of	temperature. The low-temperature data is fitted by different models. (b) Magnetoresistance (MR) of CrRhAs at selected temperatures. The gray solid line is a $\sim$B$^\alpha$ fit to the data at 2~K. The inset shows the MR values under 7~T at different temperatures. (c) Temperature dependent Hall coefficients obtained from fitting the Hall resistivity $\rho_{xy}$ shown in the inset.} \label{Fig4}
\end{figure*}

\begin{figure}
	\includegraphics[width=7.8cm]{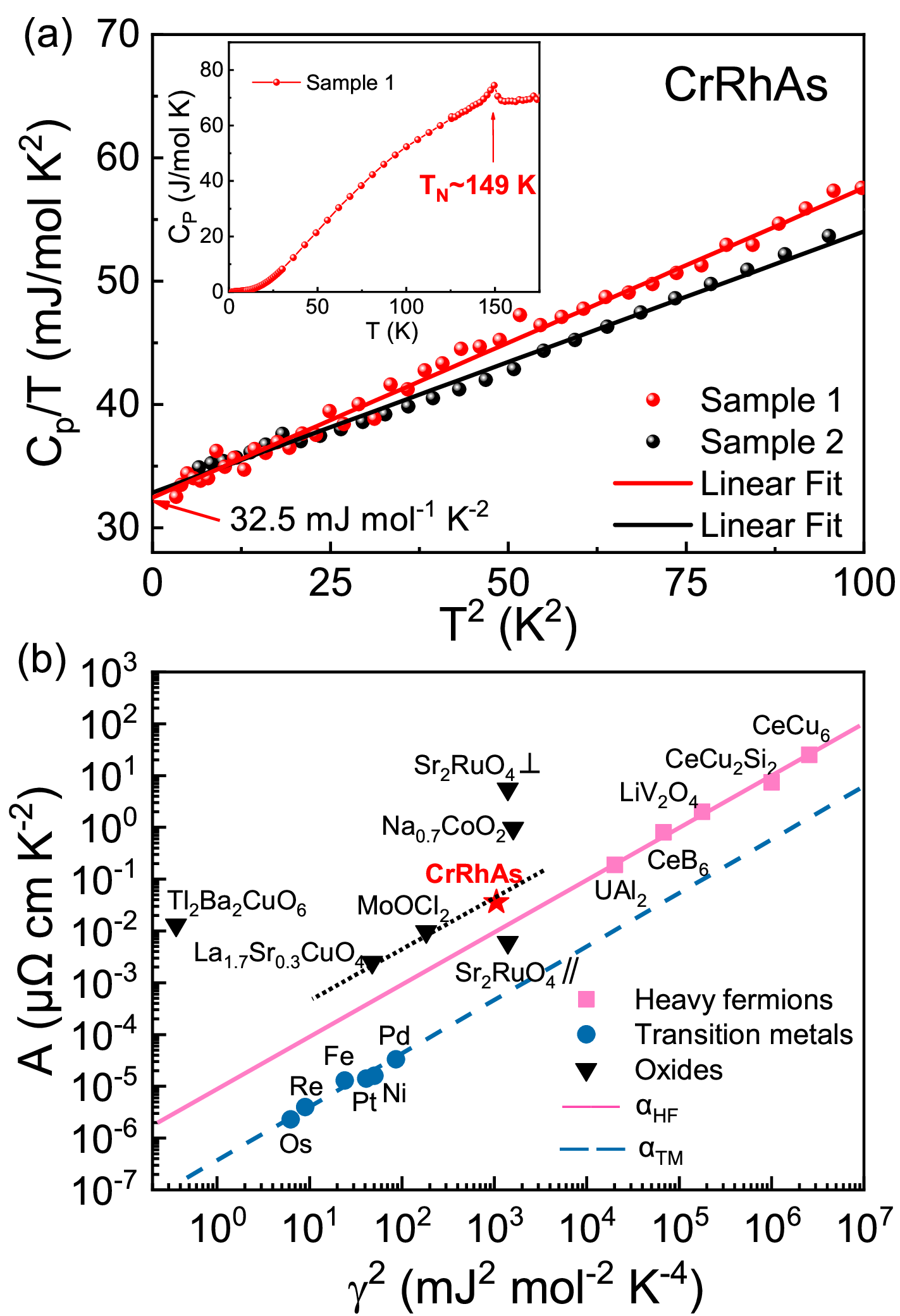}
	\caption {(a) Specific heat C$_P$/T as a function of T$^2$ at low-temperature region for two different samples is plot. The inset shows the temperature-dependent specific heat C$_P$ of CrRhAs. (b) Kadowaki-Woods plot of A versus $\gamma^2$ in a	log-log scale for different classes of materials.} \label{Fig5}
\end{figure}

\subsection{Transport properties and heat capacity}

The electrical transport properties of CrRhAs polycrystalline sample are presented in Fig.\ref{Fig4}. From room temperature, the resistivity $\rho$(T) first increases with cooling then drops drastically below T$_N$. The semiconducting-like behavior above T$_N$ may be due to the strong spin fluctuations or short-range magnetic order, which localize the charge carriers\cite{EuZn2As2_TologicalHall_PRB}. When the long-range magnetic order well establishes below T$_N$, the metallic behavior emerges. Assuming the drop of $\rho$(T) is mainly governed by the changes in the phonon scattering of the conduction electrons with decreasing temperature, the data could be fitted by the Bloch-Gr$\ddot{u}$neisen-Mott (BGM) formula\cite{TmPtIn_BGM_JMM}: 

\begin{equation*}
     \rho(T)_{BGM} = \rho_{0} + aT \left(\frac {T} {T_{D}} \right)^4 \int_0^{\frac {T_D} {T}}  \frac {x^5 dx}{ (e^x-1)(1-e^x)} - \alpha T^3
\end{equation*}

where $\rho_{0}$ is the residual resistivity and the other terms account for electron-phonon scattering processes and $s-d$ interband scattering. The fitting result is presented as the red solid line in Fig.\ref{Fig4}(a) with the following set of parameters: $\rho_{0}=429.8(2)$~$\mu\Omega$~cm, $a=8.23(1)$~$\mu\Omega$~cm/K, $T_{D}=~109.5(9)$~K and $\alpha=2.11(2)$$\times$10$^{-5}$~$\mu\Omega$~cm/K$^{3}$. Among them, T$_D$ can be viewed as an estimation of Debye temperature and reaches an agreement with the values obtained from the following specific heat data. On the other hand, the resistivity data below 30~K can be well fitted by the formula $\rho(T)=\rho_{0}+AT^2$ with the parameters: $\rho_{0}=427.4(1)$~$\mu\Omega$~cm, $A=0.0358(2)$~$\mu\Omega$~cm/K$^{^2}$. The dominant $\sim$T$^2$ dependence indicates either electron-electron or electron-magnon interactions.    

Fig.\ref{Fig4}(b) presents the magnetoresistance (MR) data and MR is defined as [$\rho$(B) - $\rho$(B = 0)]/$\rho$(B = 0). Overall, MR is quite weak at all temperatures and its maximum value is lower than 0.4\% at 14~T and 2~K. The main feature is that MR is negative above T$_N$ and positive below T$_N$. The weak negative MR above T$_N$ might result from the reduced spin scattering by applying magnetic field\cite{FeGeSb,FePdTe}. In the magnetic ordered state, the weak positive MR could be simply due to the ordinary Lorentz force-induced scattering\cite{MagnetoresistanceInMetals}. Fitting MR data at 2~K using a $\sim$B$^\alpha$ term yields the $\alpha$ value very close to 2. Such a B$^2$ dependence of MR can also be given by the two-band theory\cite{ElectronsPhononsTheory}.

Next, Hall effect measurements on CrRhAs show a remarkable amount of variation with temperature. The inset of Fig.\ref{Fig4}(c) display the Hall resistivity $\rho_{xy}$ at selected temperatures. $\rho_{xy}$ follows quasi-linear field dependence within the instrumental resolution but its slope changes from positive at low temperatures to negative above 75~K then back to positive above room temperature. Through linear fits of $\rho_{xy}$, the Hall coefficients R$_H$ at different temperatures were obtained and presented in Fig.\ref{Fig4}(c). Such behavior resembles that in the kagome metal AV$_3$Sb$_5$ and other similar systems\cite{RbV3Sb5_HallCoefficientChange_CPL,LuRuGe_HallCoefficientChange_IC,PtSn4_HallCoefficientChange_PRB,TMTMAs_PHDThesis}. The strong temperature dependence and sign reversals of R$_H$ suggest that CrRhAs is a multi-band metal, consistent with previous theoretical calculations of its band structures\cite{CrRhAs_Theory_NPJ}.

Finally, the specific heat data of CrRhAs is shown in Fig.\ref{Fig5}(a). A lamda-shape peak appears at T$_N$ confirming the antiferromagnetic transition. Plots of C$_p$/T versus T$^2$ are shown for two samples from different batches. A linear fitting at low temperatures by C$_p$/T = $\beta$T$^2$ + $\gamma$ is displayed by the solid line. The obtained parameters are roughly the same ($\beta=0.2532$~mJ mol$^{-1}$ K$^{-4}$ and $\gamma=32.5$ mJ mol$^{-1}$ K$^{-2}$ for sample 1, $\beta=0.2117$ mJ mol$^{-1}$ K$^{-2}$ and $\gamma$ = 32.8mJ mol$^{-1}$ K$^{-2}$ for sample 2). The Debye temperature is estimated to be 136.7~K for sample 1 and 145.2~K for sample 2. The obtained electronic specific heat coefficient $\gamma$ up to 32.8 mJ mol$^{-1}$ K$^{-2}$ is a very large value comparing with conventional metal.

Both $\gamma$ and the coefficient A in the scaling relation for resistivity contain the effects of electron-electron interaction. The Kadowaki-Woods ratio defined as $\alpha$ = A/$\gamma$$^2$, is an important parameter which probes the relationship between the electron-electron scattering rate and the renormalization of effective mass\cite{ModiKWR_NP}. For CrRhAs, its Kadowaki-Woods ratio is as large as 33.9 $\mu$$\Omega$ cm mol$^2$~K$^2$/J$^{2}$, which indicates that it is a strongly correlated metal. Further discussions will be made in the following section.

\subsection{Discussion}

It is important to compare the experimentally determined antiferromagnetic structure of CrRhAs in this work with that obtained from previous DFT calculations\cite{CrRhAs_Theory_NPJ}. Firstly, they have reached agreements in many aspects. For examples, both works have shown that CrRhAs has a noncollinear magnetic structure with magnetic moments lying on the $ab$-planes and the interlayer magnetic interaction is antiferromagnetic. However, there are still some major differences. Previous DFT calculations have shown that CrRhAs is dominated by an antiferromagnetic second nearest neighbor coupling $J_2$ $\sim$ 44.8~meV in the kagome plane while the nearest neighbor coupling $J_1$ $\sim$ -3.5~meV is ferromagnetic and much smaller\cite{CrRhAs_Theory_NPJ}. In contrast, our neutron diffraction result gives a clear ferromagnetic spin configuration for the second nearest neighbor coupling as shown by the blue triangles in Fig.\ref{Fig3}(e). On the other hand, by extending the magnetic susceptibility measurement to 400~K, a reasonable Curie-Weiss fit gives a CW temperature $\theta_{CW}$ = -479~K which supports a dominant antiferromagnetic interaction in this compound. Therefore based on the current experimental result, it is more likely that the magnetic ground state of CrRhAs is dominated by strong antiferromagnetic nearest neighbor coupling $J_1$ and ferromagnetic nearest neighbor coupling $J_2$ that is not very strong. Previous DFT calculations also reveal that CrRhAs has an unusual magnetic Hamiltonion that simple Heisenberg exchange
interactions are not sufficient and adding ring
exchange terms is indispensable\cite{CrRhAs_Theory_NPJ}. Although currently there is a discrepancy between the experimentally and theoretically determined magnetic structure, this also means that intriguing physics would be expected in the future work in reconciling the experimental and theoretical results.

On the other hand, CrRhAs also features a strong electron correlation supported by the large Kadowaki-Woods ratio $\alpha$. As shown in Fig.\ref{Fig5}(b), typically $\alpha_{TM}$ $\approx$ 0.4 $\mu$$\Omega$~cm~mol$^2$~K$^2$/J$^{2}$ for simple transitional metals like Fe and Ni. While $\alpha_{HF}$ reaches a large value of 10 $\mu$$\Omega$~cm~mol$^2$~K$^2$/J$^{2}$ for many heavy fermion compounds. Many famous strongly correlated oxides could take even larger values as shown by the black triangles in Fig.\ref{Fig5}(b). For CrRhAs, $\alpha$ is more than three times larger than that for typical heavy fermion compounds\cite{ModiKWR_NP}. The Kadowaki-Woods ratio usually takes similar value for one class of materials as demonstrated by the straight scaling lines in Fig.\ref{Fig5}(b). Interestingly, in the A versus $\gamma^2$ plot, CrRhAs could roughly be scaled with cuprate superconductor La$_{1.7}$Sr$_{0.3}$CuO$_{4}$ and a strongly correlated metal MoOCl$_2$ that discovered recently\cite{MoOCl2_StrongCorrelation_PRM,MoOCl2Science}. There has been some reports about heavy-fermion behavior in Yb-based ZrNiAl-type compounds\cite{YbHF1,YbHF2}, while it seems that CrRhAs is the first experimentally proposed strongly correlated metal in $3d$-transitional metal based ZrNiAl-type material. It is also interesting to mention that MnRuAs with similar crystal structure was theoretical proposed to have strong electron correlation and flat Fermi surface sheets that potentially hold density-wave like transitions\cite{MnRuAs_StrongCorrelation_PRB}. Therefore, these compounds are promising in further exploring emergent phenomena that related to strongly correlated electrons.

\section{Conclusions}

To summarize, both polycrystalline and micro-single crystal samples of CrRhAs with distorted kagome lattice were synthesized. CrRhAs is an antiferromagnet with T$_N$ $\sim$ 149~K and strong antiferromagnetic spin fluctuations up to room temperature. Neutron diffraction reveal that CrRhAs has an in-plane noncollinear antiferromagnetic structure with propagation vector $k=(1/3,1/3,1/2)$, in contrast to the magnetic structure proposed by DFT calculations previously. CrRhAs also features sign-reversals of Hall coefficients related to the multi-band effect and strong electron correlations supported by the large Kadowaki-Woods ratio. Therefore, our findings not only provide a new material platform for studying the interplay between spin frustration, multi-band effect and electron correlation, but also may stimulate new attentions on other $3d$-transitional metal based ZrNiAl-type kagome metals that has been rarely explored.

\section*{Acknowledgement}
This work was supported by the National Natural Science Foundation of China (No. 12474148 and No. 12074426)

\bibliography{CrRhAs}{}
\end{document}